\documentclass[fleqn,10pt]{wlscirep}
\usepackage[T1]{fontenc}  
\usepackage{multirow}
\usepackage{epsfig,epsf,psfrag} 
\usepackage{graphicx} 
\usepackage{subfigure}
\usepackage{amsmath}
\usepackage{hyperref}
\usepackage{pslatex}
\usepackage{epic,eepic} 
\usepackage{color,pstcol}
\usepackage{pstricks} 
\usepackage{fancyhdr}
\usepackage{sidecap}
\usepackage{mathtext}
\usepackage{inputenc}
\usepackage{babel}
\usepackage{xspace}
\title{Yu-Shiba-Rusinov bound states induced by a spin flipper in the vicinity of a s-wave superconductor} 
  \author[1]{Subhajit Pal}
\author[1,*]{Colin Benjamin}

\affil[1]{School of Physical Sciences, National Institute of Science Education \& Research, HBNI, Jatni-752050,\ India}

\affil[*]{colin.nano@gmail.com}

\keywords{Josephson junction, High spin states, Magnetic impurity}

\begin{abstract}
We theoretically study the formation and characteristics of Yu-Shiba-Rusinov bound states within the superconducting gap using a BTK approach in presence of a spin flipper (high spin magnetic impurity). We focus on the zero energy in the conductance spectra and show how a peak is formed at $E=0$ due to flipping of the magnetic impurity spin, but for no flip case a dip forms at $E=0$ in the conductance spectra. This $E=0$ conductance peak is almost quantized at $2e^2/h$ values, however it arises due to non-topological reasons in contrast to the $E=0$ peak formed due to Majorana states.    
\end{abstract}
\begin{document}
\flushbottom
\maketitle
\section*{Introduction}
In conventional s-wave superconductors magnetic impurities induce bound states, whose energy lies within the superconducting gap. This was first discovered by Yu, Shiba and Rusinov independently in the late 1960s and is now termed as Yu-Shiba-Rusinov\cite{Yu,Shib,Rusi} (YSR) states. The interaction of the impurity spin with the Andreev reflected electrons or holes gives rise to these low-lying YSR excited states. In past years YSR bound states have been observed experimentally by scanning tunneling spectroscopy\cite{yaz,frank,Ruby} on superconducting Pb or Nb surface. Further in Ref.~[\cite{ruby}] the authors consider a system where iron (Fe) chains are doped on the superconducting Pb surface. They investigate the subgap spectra using scanning tunneling microscopy.\par
In this work we show the occurrence of YSR states using a simple BTK approach\cite{BTK}, perhaps for the first time, in presence of a spin flipper or high spin magnetic impurity (HSM). The exact setting we will use is shown in Fig.~1, it depicts a HSM at $x=0$ and a $\delta$-like potential barrier at $x=a$. In the regions I ($x<0$) and II ($0<x<a$) there are two  normal metals while for $x>a$ there is a s-wave superconductor. We study the signature of YSR bound states through the Andreev reflection probabilities and conductance spectra. We see that when HSM does not flip there is a dip at zero energy ($E=0$) in the conductance spectra, but for spin flip case we get a zero energy peak due to YSR states. Technically YSR bound states are obtained by taking exchange interaction $J\rightarrow0$, impurity spin $S\rightarrow \infty$, and thus rendering $JS$=finite\cite{Shib,koer}. In our work too we see YSR states arise for low values of $J$ and high $S$ values.\par
The importance of YSR bound states was recently enhanced for several reasons. One reason is that the prediction of topological superconductivity and Majorana bound states in chains of magnetic adatoms on superconductors\cite{majo}. Majorana bound states are in high demand as the potential building blocks of a future topological quantum computer\cite{nayak}. Another reason is that there has been experimental progress on the measurement of subgap spectra with much higher resolution than previously possible. This has motivated theoretical and experimental work examining the basic properties of YSR bound states in more detail. Further, YSR bound states carry information on the strength of exchange coupling between impurity spin with the Andreev reflected electrons or holes, which measures the many-body ground state properties of the system\cite{ben}.\par  
Our paper is structured as follows: in the next section on Theory, we first introduce our model and give a background to our study by providing the Hamiltonian, wavefunctions and boundary conditions used to calculate the different reflection probabilities. We show the signature of YSR bound states in the normal and  Andreev reflection probability, and differential conductance in the succeeding section. The section after that deals with Yu-Shiba-Rusinov bound states, wherein we focus on the signature of Yu-Shiba-Rusinov states in the conductance spectra. Finally, we give a brief conclusion to our study.  
\section*{Theory}
\subsection*{Hamiltonian}
The Hamiltonian\cite{AJP,mauri} used to describe a HSM is given by-
\begin{equation}
H_{HSM}=-J_{0}\vec{s}.\vec{S}  
\end{equation}
The above model for a magnetic impurity matches quite well with solid-state scenarios such as seen in  1D quantum wires or graphene  with an embedded magnetic impurity or quantum dot\cite{palma,zac}. 
Electrons interact with magnetic impurity via $-J_{0}\vec{s}.\vec{S}$, where $J_{0}$ being the strength of the exchange interaction, $\vec{s}$ is the electronic spin and $\vec{S}$ is the spin of the magnetic impurity. $J_{0}(=\frac{\hbar^2 k_{F}J}{m^{\star}})$, with $J$ being the relative magnitude of the exchange interaction, $m^{\star}$ is the electronic mass and Fermi wavevector $k_{F}$ is defined via the Fermi energy $E_{F}$ which is the largest energy scale in our system (around~$1000\Delta$), $\Delta$ being the superconducting gap which for a widely used s-wave superconductor like Aluminium is around $0.17$ meV.\par
In our work we consider a metal (N)-metal (N)-superconductor (S) junction with a HSM between two metals at ($x=0$) and a $\delta$-like potential barrier exists at metal-superconductor interface at ($x=a$). When an electron with energy E and spin ($\uparrow$) is incident from the normal metal, at the $x=0$ interface it interacts with the HSM through an exchange potential which may induce a mutual spin flip. The electron can be reflected back to region I, or transmitted to region II, with spin up or down. When this transmitted electron is incident at $x=a$ interface it could be reflected back from the interface and there is also the possibility of Andreev reflection, i.e., a hole with spin up or down is reflected back to region II. Electron-like and hole-like quasi-particles with spin up or down are transmitted into the superconductor for energies above the gap.

The model Hamiltonian in Bogoliubov-de Gennes formalism of our Normal Metal-Magnetic impurity-Normal Metal-Insulator-Superconductor system is given below:
\begin{eqnarray}
\begin{bmatrix}
H\hat{I} & i\Delta \theta(x-a)\hat{\sigma}_{y}  \\
-i\Delta^{*}\theta(x-a)\hat{\sigma}_{y}   &  -H\hat{I}
\end{bmatrix} \psi(x)& =& E \psi(x), \mbox{ where } H = p^2/2m^\star + V\delta(x-a) - J_{0}\delta(x)\vec s.\vec S -E_{F},\nonumber\\
&&\mbox{ $\psi$ is a four-component spinor, $\Delta$ is the gap in s-wave superconductor and }\nonumber\\
&&\theta \mbox{   is the Heaviside step function.}
\end{eqnarray}
Further, in $H$ the first term is the kinetic energy of an electron with effective mass $m^\star$, $V$ is the strength of the $\delta$-like potential at the interface between normal metal and superconductor, the third term describes the exchange interaction (of strength $J_{0}$) between the electron with spin $\vec s$ and high spin magnetic impurity with spin $\vec S$, $\hat{\sigma}$ is the Pauli spin matrix and $\hat{I}$ is the identity matrix, $E_F$ being the Fermi energy. We will later use the dimensionless parameter $J=\frac{m^{\star}J_{0}}{\hbar^2 k_{F}}$ as a measure of strength of exchange interaction\cite{AJP} and $Z=\frac{m^{\star}V}{\hbar^2 k_{F}}$ as a measure of interface transparency\cite{BTK}. In our work $Z$ is a dimensionless quantity, while $V$ has the dimensions of energy. $Z$ denotes the transparency of the junction, $Z=0$ means completely transparent junction, while $Z>>1$ implies a tunneling junction\cite{BTK,b}.\\
\begin{figure}[h]
\centering{\includegraphics[width=0.8\linewidth]{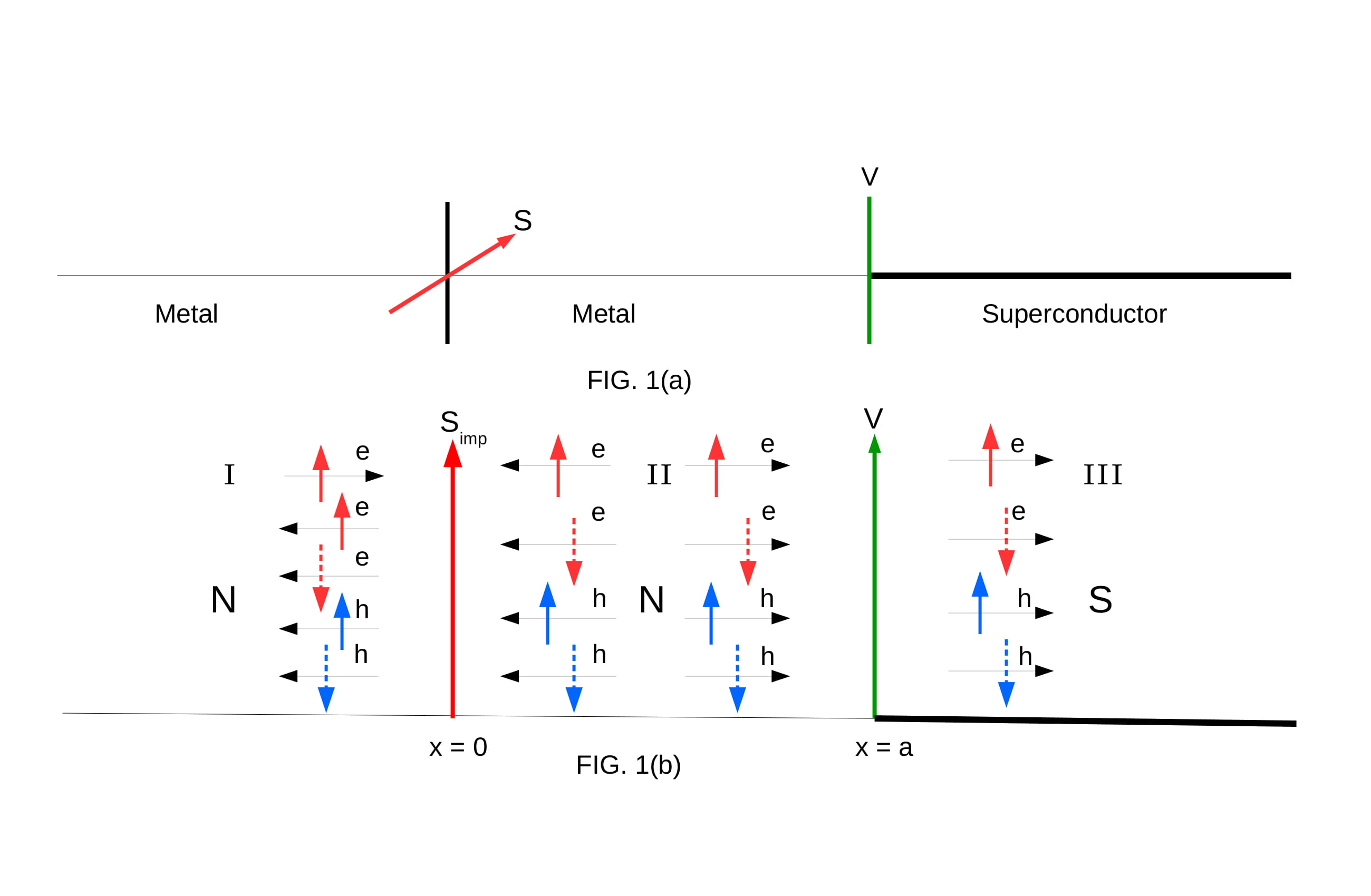}}
\caption{(a) A high spin magnetic impurity with spin S and magnetic moment $m'$ at $x=0$ in a Normal Metal-HSM-Normal Metal-Insulator-Superconductor (NMNIS) junction, (b) The scattering of an up-spin electron incident is shown. Andreev reflection and quasi particle transmission into superconductor are depicted. For details see Eqs.~3,4,5.}
\end{figure}
\subsection*{Wavefunctions}
The wavefunctions in the different regions of the system are as shown in Figs. 1(a) and 1(b) and can be written in spinorial form\cite{LINDER} for an electron with spin up incident from region I (normal metal) as:
\begin{equation}
\psi_{N}^{I}(x)=\begin{pmatrix}
                    1\\
                    0\\
                    0\\
                    0
                  \end{pmatrix}e^{ik_{e}x}\phi_{m'}^{S}+r_{ee}^{\uparrow\uparrow}\begin{pmatrix}
                  1\\
                  0\\
                  0\\
                  0
                 \end{pmatrix}e^{-ik_{e}x}\phi_{m'}^{S}+r_{ee}^{\uparrow\downarrow}\begin{pmatrix}
                 0\\
                 1\\
                 0\\
                 0
                \end{pmatrix}e^{-ik_{e}x}\phi_{m'+1}^{S}+r_{eh}^{\uparrow\uparrow}\begin{pmatrix}
                0\\
                0\\
                1\\
                0
               \end{pmatrix}e^{ik_{h}x}\phi_{m'+1}^{S}+r_{eh}^{\uparrow\downarrow}\begin{pmatrix}
               0\\
               0\\
               0\\
               1
              \end{pmatrix}e^{ik_{h}x}\phi_{m'}^{S},\mbox{for $x<0,$}
\end{equation}              
\begin{eqnarray}
\psi_{N}^{II}(x)=t_{ee}^{'\uparrow\uparrow}\begin{pmatrix}
                                     1\\
                                     0\\
                                     0\\
                                     0
                                     \end{pmatrix}e^{ik_{e}x}\phi_{m'}^{S}+t_{ee}^{'\uparrow\downarrow}\begin{pmatrix}
                                     0\\
                                     1\\
                                     0\\
                                     0
                                    \end{pmatrix}e^{ik_{e}x}\phi_{m'+1}^{S}+b_{ee}^{\uparrow\uparrow}\begin{pmatrix}
                                    1\\
                                    0\\
                                    0\\
                                    0
                                   \end{pmatrix}e^{-ik_{e}(x-a)}\phi_{m'}^{S}+b_{ee}^{\uparrow\downarrow}\begin{pmatrix}
                                   0\\
                                   1\\
                                   0\\
                                   0
                                  \end{pmatrix}e^{-ik_{e}(x-a)}\phi_{m'+1}^{S}\nonumber\\+c_{eh}^{\uparrow\uparrow}\begin{pmatrix}
                                  0\\
                                  0\\
                                  1\\
                                  0
                                 \end{pmatrix}e^{ik_{h}(x-a)}\phi_{m'+1}^{S}+c_{eh}^{\uparrow\downarrow}\begin{pmatrix}
                                 0\\
                                 0\\
                                 0\\
                                 1
                                \end{pmatrix}e^{ik_{h}(x-a)}\phi_{m'}^{S}+a_{eh}^{\uparrow\uparrow}\begin{pmatrix}
                                0\\
                                0\\
                                1\\
                                0
                               \end{pmatrix}e^{-ik_{h}x}\phi_{m'+1}^{S}+a_{eh}^{\uparrow\downarrow}\begin{pmatrix}
                               0\\
                               0\\
                               0\\
                               1
                              \end{pmatrix}e^{-ik_{h}x}\phi_{m'}^{S},\mbox{for $0<x<a,$}
                              \end{eqnarray}
\begin{equation}                              
\mbox{ and }\psi_{S}(x)=t_{ee}^{\uparrow\uparrow}\begin{pmatrix}
                              u\\
                              0\\
                              0\\
                              v
                             \end{pmatrix}e^{iq_{+}x}\phi_{m'}^{S}+t_{ee}^{\uparrow\downarrow}\begin{pmatrix}
                             0\\
                             u\\
                             -v\\
                             0
                             \end{pmatrix}e^{iq_{+}x}\phi_{m'+1}^{S}+t_{eh}^{\uparrow\uparrow}\begin{pmatrix}
                             0\\
                             -v\\
                             u\\
                             0
                             \end{pmatrix}e^{-iq_{-}x}\phi_{m'+1}^{S}+t_{eh}^{\uparrow\downarrow}\begin{pmatrix}
                             v\\
                             0\\
                             0\\
                             u
                             \end{pmatrix}e^{-iq_{-}x}\phi_{m'}^{S},\mbox{for $x>a.$}
\end{equation}                             
$r_{ee}^{\uparrow\uparrow}$($r_{ee}^{\uparrow\downarrow}$) and $r_{eh}^{\uparrow\uparrow}$($r_{eh}^{\uparrow\downarrow}$) are the corresponding  amplitudes for normal reflection and Andreev reflection with spin up(down). $t_{ee}^{\uparrow\uparrow}$($t_{ee}^{\uparrow\downarrow}$) and $t_{eh}^{\uparrow\uparrow}$($t_{eh}^{\uparrow\downarrow}$) are the corresponding amplitudes for transmission of electron-like quasi-particles and hole-like quasi-particles with spin up(down). $\phi_{m'}^{S}$ is the eigenfunction of magnetic impurity: with its $S^{z}$ operator acting as-
$S^{z}\phi_{m'}^{S} = m'\phi_{m'}^{S}$, with $m'$ being the spin magnetic moment of the HSM.  
For $E>\Delta$ (for energies above the gap), the BCS coherence factors are $u=\sqrt{\frac{1}{2}\Big[1+\frac{(E^2-\Delta^2)^\frac{1}{2}}{E}\Big]}$, $v=\sqrt{\frac{1}{2}\Big[1-\frac{(E^2-\Delta^2)^\frac{1}{2}}{E}\Big]}$, while the wave-vector in metal is $k_{e,h}=\sqrt{2m^{\star}(E_{F}\pm E)}$ and in superconductor is $q_{\pm}=\sqrt{2m^{\star}(E_{F}\pm \sqrt{E^2-\Delta^2})}$ and for $E<\Delta$ (for energies below the gap) the BCS coherence factors are $u=\sqrt{\frac{1}{2}\Big[\frac{E+i(\Delta^2-E^2)^\frac{1}{2}}{\Delta}\Big]}$, $v=\sqrt{\frac{1}{2}\Big[\frac{E-i(\Delta^2-E^2)^\frac{1}{2}}{\Delta}\Big]}$, while the wave-vector in metal remains same, and in superconductor is $q_{\pm}=\sqrt{2m^{\star}(E_{F}\pm i\sqrt{\Delta^2-E^2})}$\cite{BTK}, wherein $E$ is the excitation energy of electron above $E_{F}$. In Andreev approximation, which we will use throughout this work, $E_{F}\gg \Delta, E$, thus we have $k_{e}=k_{h}=q_{+}=q_{-}=k_{F}$.
\subsection*{Boundary conditions}
The boundary condition at $x=0$ is- 
\begin{eqnarray}
{}&\psi_{N}^{I}(x)=\psi_{N}^{II}(x)\mbox{ (continuity of wavefunctions), }\\ 
&\mbox { and, }\frac{d\psi_{N}^{II}}{dx}-\frac{d\psi_{N}^{I}}{dx}=-\frac{2m^{\star}J_{0}\vec s.\vec S}{\hbar^2} \psi_{N}^{I}\mbox{ (discontinuity in first derivative) }, 
\end{eqnarray}
$\vec{s}.\vec{S}$ is the exchange operator in the Hamiltonian and is given by\cite{AJP} $\vec s.\vec S=s^{z}S^{z}+\frac{1}{2}(s^{-}S^{+}+s^{+}S^{-})$, $s^{\pm}=s_{x}\pm is_{y}$ is the raising and lowering operator for electron and $S^{\pm} = S_{x}\pm iS_{y}$ is the raising and lowering spin operator for HSM, and boundary condition at $x=a$ is-
\begin{eqnarray}
{}&\psi_{N}^{II}(x)=\psi_{S}(x)\mbox{ (continuity of wavefunctions), }\\  
&\mbox{ and, }\frac{d\psi_{S}}{dx}-\frac{d\psi_{N}^{II}}{dx}=\frac{2m^{\star}V}{\hbar^2}\psi_{N}^{II}\mbox{ (discontinuity in first derivative). } 
\end{eqnarray}
When an electron with spin up is incident from the metallic region, at the $x=0$ interface it interacts with the HSM via the exchange operator $\vec{s}.\vec{S}$ in the Hamiltonian which may induce a mutual spin flip. The electron can be reflected back to region I with spin up or down and in presence of superconductor there is also possibility of Andreev reflection, i.e., a hole with spin up or down is reflected back to region I (see Fig.~1(b)). Thus the wavefunction of the first metallic region $\psi_N^I$ (Eq.~3) has four components with (a) spin up electron, (b) spin down electron, (c) spin up hole, and (d) spin down hole. Now when the exchange operator $\vec{s}.\vec{S}$ acts on the wavefunction $\psi_N^I$ via Eq.~7 we get-\\  
For spin up electron component:
\begin{equation}
 \vec s.\vec S\begin{pmatrix}
               1\\
               0\\
               0\\
               0
              \end{pmatrix}\phi_{m'}^{S}=\frac{m'}{2}\begin{pmatrix}
              1\\
              0\\
              0\\
              0
             \end{pmatrix}\phi_{m'}^{S}+\frac{F}{2}\begin{pmatrix}
             0\\
             1\\
             0\\
             0
            \end{pmatrix}\phi_{m'+1}^{S}\mbox{, where $F=$ spin flip probability of HSM$=\sqrt{(S-m')(S+m'+1)}$,}\nonumber
\end{equation}
while for spin down electron component:
\begin{equation}
\vec s.\vec S\begin{pmatrix}
              0\\
              1\\
              0\\
              0\\
             \end{pmatrix}\phi_{m'+1}^{S}=-\frac{(m'+1)}{2}\begin{pmatrix}
             0\\
             1\\
             0\\
             0
             \end{pmatrix}\phi_{m'+1}^{S}+\frac{F}{2}\begin{pmatrix}
                                                                1\\
                                                                0\\
                                                                0\\
                                                                0
                                                               \end{pmatrix}\phi_{m'}^{S},\nonumber
\end{equation}
for spin up hole component:
\begin{equation}
\vec s.\vec S\begin{pmatrix}
              0\\
              0\\
              1\\
              0
             \end{pmatrix}\phi_{m'+1}^{S}=-\frac{(m'+1)}{2}\begin{pmatrix}
             0\\
             0\\
             1\\
             0
             \end{pmatrix}\phi_{m'+1}^{S}+\frac{F}{2}\begin{pmatrix}
             0\\
             0\\
             0\\
             1
             \end{pmatrix}\phi_{m'}^{S},\nonumber
\end{equation}
and finally for spin down hole component:
\begin{equation}
\vec s.\vec S\begin{pmatrix}
              0\\
              0\\
              0\\
              1
             \end{pmatrix}\phi_{m'}^{S}=\frac{m'}{2}\begin{pmatrix}
             0\\
             0\\
             0\\
             1
            \end{pmatrix}\phi_{m'}^{S}+\frac{F}{2}\begin{pmatrix}
            0\\
            0\\
            1\\
            0
            \end{pmatrix}\phi_{m'+1}^{S}.\nonumber
\end{equation}
Here $F=\sqrt{(S-m')(S+m'+1)}$ is the spin-flip probability\cite{AJP} for HSM. Using the above equations and from boundary conditions (Eqs.~(6-9)) we get 16 equations. We solve the 16 equations to calculate the different normal and Andreev reflection probabilities: $R_{ee}^{\uparrow\uparrow}= |r_{ee}^{\uparrow\uparrow}|^{2},R_{ee}^{\uparrow\downarrow}=|r_{ee}^{\uparrow\downarrow}|^{2},R_{eh}^{\uparrow\uparrow}=|r_{eh}^{\uparrow\uparrow}|^{2},R_{eh}^{\uparrow\downarrow}=|r_{eh}^{\uparrow\downarrow}|^{2}$.
\subsection*{Andreev and normal reflection probability}
In Fig.~2 we plot the normal and Andreev reflection probabilities with spin flip or no flip for different values of the spin of magnetic impurity $S$ $(19/2, 21/2, 23/2)$, we fix the magnetic moment of the magnetic impurity- $m'=-1/2$ and we take $Z=0.85$- the non transparent regime. In Fig.~2(a) we plot the normal reflection probability without spin flip for both below as well as above the gap. We see that at zero energy $E=0$ there is a dip, related to the band of YSR states and it is robust for high impurity spin $S=23/2$. 
\begin{figure*}  
\includegraphics[width=0.75\textwidth]{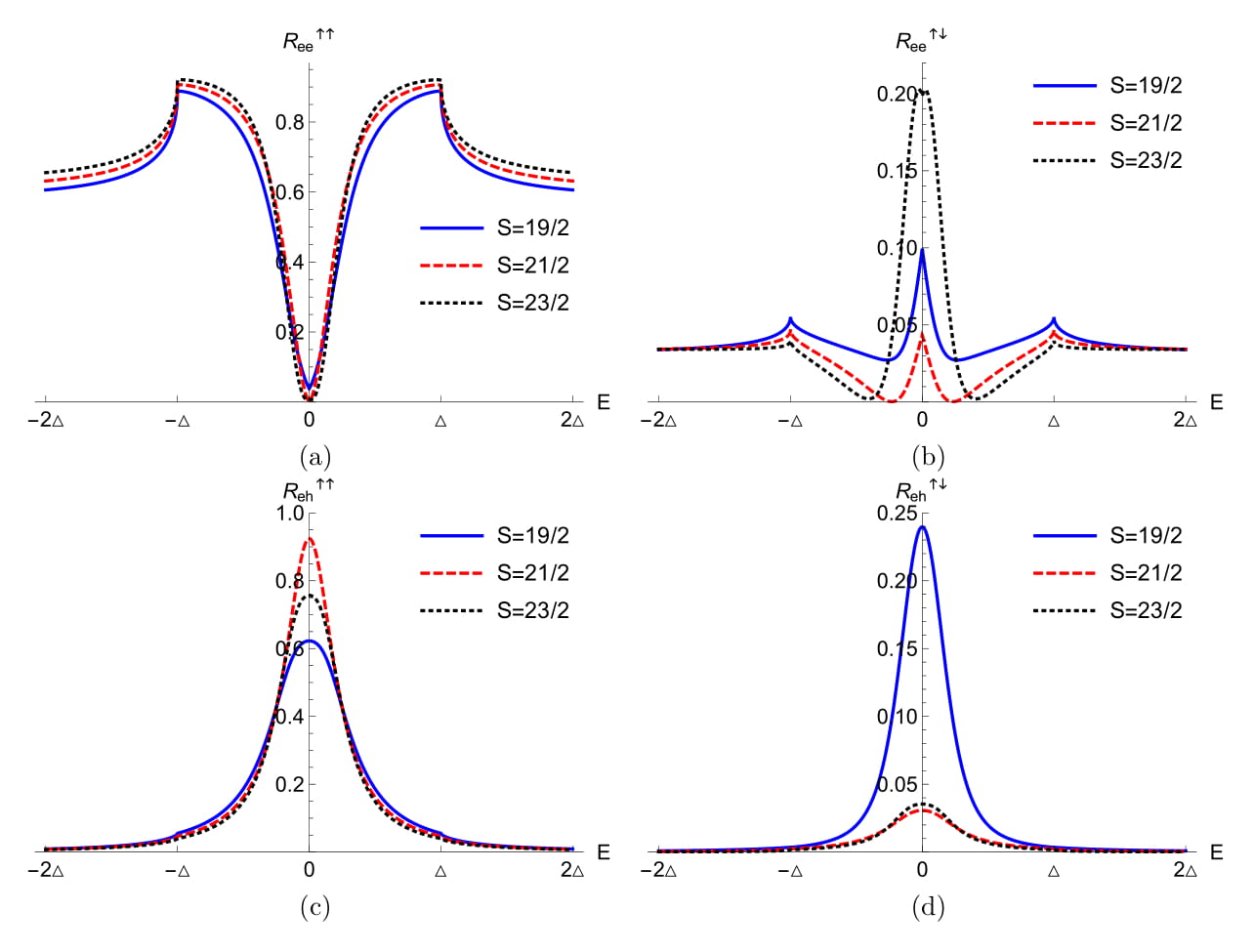}
\caption{\small \sl a) Normal reflection  probability without flip, b) Normal reflection probability with flip, c) Andreev reflection probability with flip, d) Andreev reflection probability without flip. Parameters are: $J=0.4$, $Z=0.85$, $m'=-1/2$ and $k_{F}a=0.8437\pi$.}
\end{figure*}
In Fig.~2(b) we plot the normal reflection probability with spin flip for below and above the gap. We see that there is a peak, related to the YSR bound states at zero energy $E=0$ within the energy gap. The explanation of why we address the zero energy peaks as related to YSR states is given in the next section, where we show how they are arise by plotting the real part of complex poles of the conductance.\par 
Next in Fig.~2(c) we plot the Andreev reflection probability with spin flip for both below and above the gap. We see that here also a peak appears at zero energy $E=0$ due to the YSR bound states within the energy gap.
Finally, in Fig.~2(d) we plot the Andreev reflection probability without flip for both below as well as above the gap. We see that there is a peak at $E=0$, related to the YSR bound states. In Ref.~[\cite{pers}] YSR states are also studied in normal-metal/superconductor junctions with magnetic impurities which are randomly orientated or ferromagnetically aligned on the surface of the superconductor. They also see the signature of YSR states in the normal and Andreev reflection probabilities. But in contrast to our case there is no peak or dip at zero energy ($E=0$) in the reflection probabilities.

\subsection*{Differential conductance \& Probability density}
To calculate the differential charge conductance, we follow the well established definitions as in Refs.~\cite{jin,kashiwaya}. The differential charge conductance is defined as-
\begin{equation}
G_{c}=1+R_{eh}^{\uparrow\uparrow}+R_{eh}^{\uparrow\downarrow}-R_{ee}^{\uparrow\uparrow}-R_{ee}^{\uparrow\downarrow} 
\end{equation}

where, 
$R_{eh}^{\uparrow\uparrow}$ is the probability of Andreev reflection of an electron (spin up)as hole (spin up),
$R_{eh}^{\uparrow\downarrow}$ is the probability of Andreev reflection of an electron (spin up)as hole (spin down),
$R_{ee}^{\uparrow\uparrow}$ is the probability of normal reflection of an electron (spin up)as electron (spin up),
$R_{ee}^{\uparrow\downarrow}$ is the probability of normal reflection of an electron (spin up)as electron (spin down).

After deriving the amplitudes of normal and Andreev reflection by solving the scattering problem, we get an expression for the differential charge conductance for $k_{F}a=0$ as follows-{
\begin{align}
G_{c}={}&\frac{R}{Q}, \mbox{ where }\\
\begin{split}
R={}& 8\Delta^2(4E^2+(2+J^2((1+m')^2-F^2)+4J(1+m')Z+4Z^2)^2(\Delta^2-E^2)+F^2J^2((4+(J+2Jm')^2)\Delta^2-J(1+2m')\\
        & E(J(E+2m'E)+4\sqrt{\Delta^2-E^2}))),\nonumber  
\end{split}\\\nonumber
\begin{split}
\mbox{ and } Q={}& 4E^2(4+J^2(1+2F^2+2m'+2m'^2)+4JZ+8Z^2)^2(\Delta^2-E^2)+(E^2(J^4(F^2+m'+m'^2)^2-4J^3(F^2+m'+m'^2)Z\\
                 & +8J(Z+2Z^3)+8(1+2Z^2+2Z^4)+J^2(2+4Z^2+F^2(4-8Z^2)+m'(4-8 Z^2)+m'^2(4-8Z^2)))-(J^4(F^2+m'\\
                 & +m'^2)^2-4J^3(F^2+m'+m'^2)Z+4(1+2Z^2)^2+8J(Z+2Z^3)+J^2(2+(4-8 F^2)Z^2+m'(4-8 Z^2)+m'^2(4-8Z^2)))\Delta^2)^2\nonumber
\end{split}\nonumber
\end{align}
The differential charge conductance at zero bias ($E=0$) from Eq.~11 we get-
\begin{equation}
G_{c}=\frac{8(F^2 J^2 (4+(J+2 J m')^2)+(2+J^2 (-F^2+(1+m')^2)+4 J (1+m') Z+4 Z^2)^2)}{(J^4(F^2+m'+m'^2)^2-4 J^3 (F^2+m'+m'^2) Z+4 (1+2 Z^2)^2+8J (Z+2 Z^3)+J^2 (2+(4-8 F^2) Z^2+m' (4-8 Z^2)+m'^2 (4-8 Z^2)))^2}
\end{equation}
From complex poles of the conductance $G_{c}$ in Eqs.~10, 11 one can get the YSR bound states $E^{\pm}$. Real part of the poles gives the energy where YSR peaks occur, while the imaginary part gives the width of the peak. For $k_{F}a=0$ we get-
\begin{equation}
\frac{E^{\pm}}{\Delta}=\pm\sqrt{\frac{X_{0}}{Y_{0}}}, \mbox{ where }
\end{equation}
\small{
\begin{align}
\begin{split}
X_{0}={}& (J^2(F^2+m'+m'^2)-2JZ-4Z^2) ((J^2 (F^2+m'+m'^2)-2JZ-4Z^2)^2+2(2+J^2(1+2m'(1+m'))+4JZ+8Z^2))\\
        & \sqrt{(J^2(F^2+m'+m'^2)-2JZ-4Z^2)^2+4(4+J^2(1+2F^2+2m'(1+m'))+4JZ+8Z^2)}+J^8 (F^2+m'+m'^2)^4-8 J^7\\
        & (F^2+m'+m'^2)^3Z+64(1+Z^2)(Z+2 Z^3)^2+32 J (Z+10 Z^3+24 Z^5+16 Z^7)-4 J^6 (F^2+m'+m'^2)^2 (-1-6 Z^2+(m'+\\
        & m'^2)(-2+4 Z^2) +F^2 (-1+4 Z^2))+16 J^5 (F^2+m'+m'^2) Z (-1+(-2+6 F^2) Z^2+(m'+m'^2) (-1+6 Z^2))-16 J^3 Z(-1\\
        & -8 Z^2-8 Z^4+3 F^2 (1+4 Z^2+8 Z^4)+(m'+m'^2)(1+8 Z^2+24 Z^4))-8 J^2 (-1-14 Z^2-56 Z^4 -48 Z^6+2 m'(1+m')\\
        & (-1+2 Z^2+8 Z^4+16 Z^6)+2 F^2 (1+6 Z^2+12 Z^4+16 Z^6))+2 J^4 (1+8 Z^2+8 Z^4+(4 m'^3+2m'^4)(5-8 Z^2+24 Z^4)-4m'\\ 
        & (-1+8 Z^2+24 Z^4)-2 m'^2 (-7+24 Z^2+24 Z^4)+F^4 (2+48 Z^4)+4 F^2 (-2 Z^2 (5+12 Z^2)+(m'+m'^2)(3-4 Z^2+24 Z^4))),\nonumber\\
\end{split}\\\nonumber\\\nonumber
Y_{0}={}& 2(J^2 (F^2+m'+m'^2)-2JZ-4Z^2)^2((J^2(F^2+m'+m'^2)-2JZ-4Z^2)^2+4(4+J^2(1+2F^2+2m'(1+m'))+4JZ+8Z^2)).\nonumber
\end{align}}
\normalsize{The condition when two YSR bound state energies merge at zero energy, i.e., $E^{\pm}=0$, is then from Eq.~13,}-
\footnotesize{
\begin{equation}
\frac{8Z(J+2 Z)(2 F^4 J^2+(1+2 m')^2+2 F^2 (J^2 m' (1+m')-Z(J+2Z)))}{J^2}=(4 F^6 J^2+8 F^4 J^2 m' (1+m')-(1+2 m')^2+4 F^2 (1+4 m'+(4+J^2) m'^2+2 J^2 m'^3+J^2 m'^4))
\end{equation}}}
The above condition leads to formation of the peaks at $E=0$ in the conductance spectra. In absence of HSM ($J=0$ case) and for no flip ($F_{2}=0$) there are no YSR states within gap ($-\Delta,\Delta$).
The signature of YSR bound states can also be seen in the probability density. To evaluate this we integrate the squared absolute value of wavefunction amplitude in normal metal region II.
\begin{equation}
\int_{0}^{a}|\psi_{N}^{II}(x)|^2dx=P=|t_{ee}^{'\uparrow\uparrow}|^2+|t_{ee}^{'\uparrow\downarrow}|^2+|b_{ee}^{\uparrow\uparrow}|^2+|b_{ee}^{\uparrow\downarrow}|^2+|c_{eh}^{\uparrow\uparrow}|^2+|c_{eh}^{\uparrow\downarrow}|^2+|a_{eh}^{\uparrow\uparrow}|^2+|a_{eh}^{\uparrow\downarrow}|^2 
\end{equation}
where $t_{ee}^{'\uparrow\uparrow}$, $t_{ee}^{'\uparrow\downarrow}$, $b_{ee}^{\uparrow\uparrow}$, $b_{ee}^{\uparrow\downarrow}$, $c_{eh}^{\uparrow\uparrow}$, $c_{eh}^{\uparrow\downarrow}$, $a_{eh}^{\uparrow\uparrow}$, $a_{eh}^{\uparrow\downarrow}$ are the reflection amplitudes of electrons and holes with spin up and down in region II (normal metal). Effectively, we sum the mod squared amplitudes of the various reflection amplitudes in the normal metal region II.

\begin{figure*} 
\includegraphics[width=.99\textwidth]{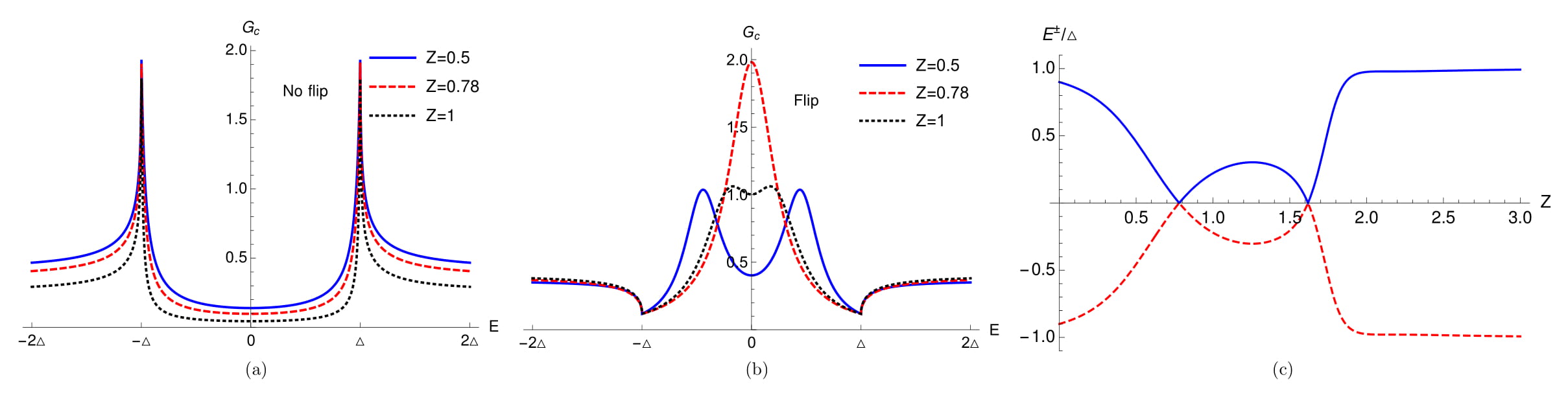}
\caption{\small \sl (a) Charge conductance vs energy for no flip case. Parameters are: $J=0.4$, $S=19/2$, $m'=19/2$, $k_{F}a=0.8437\pi$, (b) Charge conductance vs energy for spin flip case. Parameters are: $J=0.4$, $S=19/2$, $m'=-1/2$, $k_{F}a=0.8437\pi$, (c) Energy bound states as a function of interface transparency $Z$. Parameters are $S=19/2$, $m'=-1/2$, $J=0.4$, $k_{F}a=0.8437\pi$. Here charge conductance is in units of $e^2/h$.}
\end{figure*}
\section*{Yu-Shiba-Rusinov bound states}
In our work we see the YSR bound states particularly focusing on the $E=0$ YSR peak. {In Fig.~3 we plot the conductance spectra for both no flip as well as spin flip case. We see that for no flip case, there is a dip at $E=0$ for all values of interface transparency $Z$ ($Z=0.5, 0.78, 1$).
But in contrast to no flip case for a spin flip case we see that for $Z=0.78$ a peak occurs at $E=0$ due to two YSR states merging. But for $Z=0.5$ and $Z=1$, there are dips at $E=0$ like no flip case. In Fig.~3(b) we also see that there are peaks, due to the YSR bound states, present symmetrically at both positive and negative energies for $Z=0.5$ and $Z=1$. The calculated real part of poles of conductance for spin flip case in Fig.~3(b) are: $\pm 0.456033\Delta$ (for $Z=0.5$), $\pm 0.000277183\Delta$ (for $Z=0.78$), $\pm 0.221027\Delta$ (for $Z=1$) and they clearly match with the conductance peaks shown in Fig.~3(b). In Fig.~3(c) we plot energy bound state as a function of interface transparency $Z$ for the same parameters as shown in Fig.~3(b). We see that two energy bound states merge at $Z=0.78$ and $1.62$, where zero energy peaks are observed in the conductance spectra due to the YSR states.}\par
In Fig.~4(a) we plot charge conductance as a function of energy $E$ for $S=7/2$ and different values of $J$ ($J=0.3,0.4$). {We see that there is a peak, related to the YSR states, present symmetrically at both positive and negative energies for both $J=0.3$ and $J=0.4$. The energies where YSR peaks occur, i.e., values of the real part of the pole of conductance calculated from Eq.~13, i.e., $\pm 0.633701\Delta$ (for $J=0.3$) and $\pm 0.695161\Delta$ (for $J=0.4$) match quite well with the conductance peaks shown in Fig.~4(a). In Fig.~4(b) we plot charge conductance spectra for impurity spin $S=9/2$. Here we also find peaks due to YSR bound states near the gap edge within the energy gap for both $J=0.3$ and $J=0.4$. The energies where YSR peaks occur, i.e., values of the real part of the pole of conductance calculated from Eq.~13 $\pm 0.68021\Delta$ (for $J=0.3$) and $\pm 0.749387\Delta$ (for $J=0.4$) match quite well with the conductance peak shown in Fig.~4(b). In Fig.~4(c) we plot charge conductance spectra for impurity spin $S=11/2$. For $J=0.3$ and $J=0.4$ we note that there are peaks due to the YSR states near the gap edges within the energy gap.
In Figs.~4(a), (b) and (c) we give the comparison with the $J=0$ case (absence of HSM) wherein there are no YSR bound state peaks in the conductance spectra. The energies where YSR peaks occur, i.e., values of the real part of the pole of conductance calculated from Eq.~13 $\pm 0.723525\Delta$ (for $J=0.3$) and $\pm 0.794386\Delta$ (for $J=0.4$) match quite well with the conductance peaks shown in Fig.~4(c). Thus we can conclude that it is spin-flip scattering enabled by the HSM which is the reason behind occurrence of YSR bound states near the gap edges.}\par
\begin{figure*}
\includegraphics[width=0.99\textwidth]{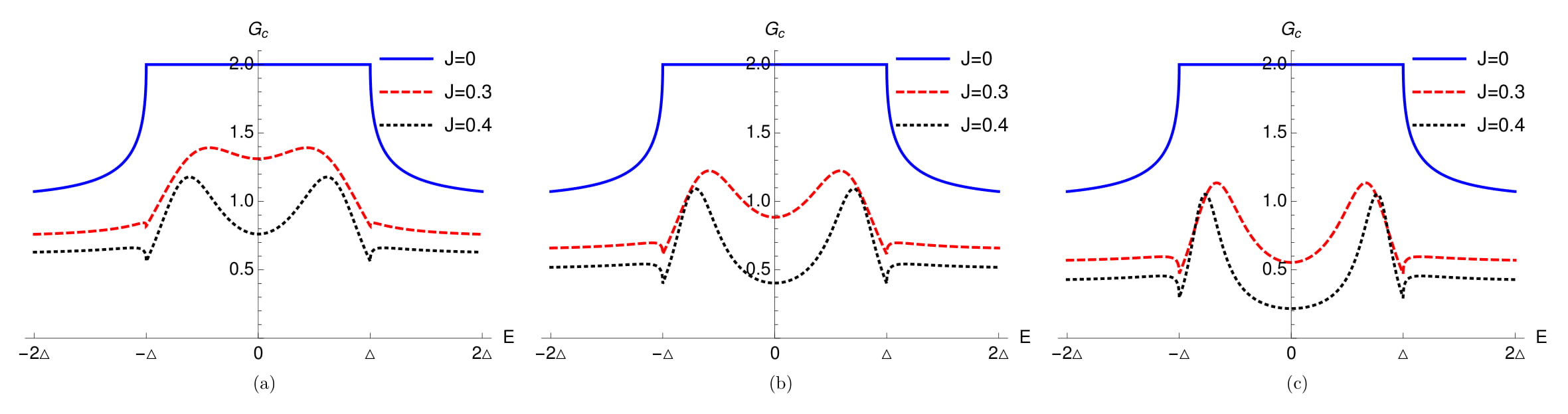}
\caption{\small\sl(a) Charge conductance vs energy in the transparent regime with $S=7/2,m'=-1/2$, (b) Charge conductance vs energy in the transparent regime with $S=9/2,m'=-1/2$, (c) Charge conductance vs energy in the transparent regime with $S=11/2, m'=-1/2$. Here charge conductance is in units of $e^2/h$.}
\end{figure*}
\begin{figure}[h]  
\includegraphics[width=.65\textwidth]{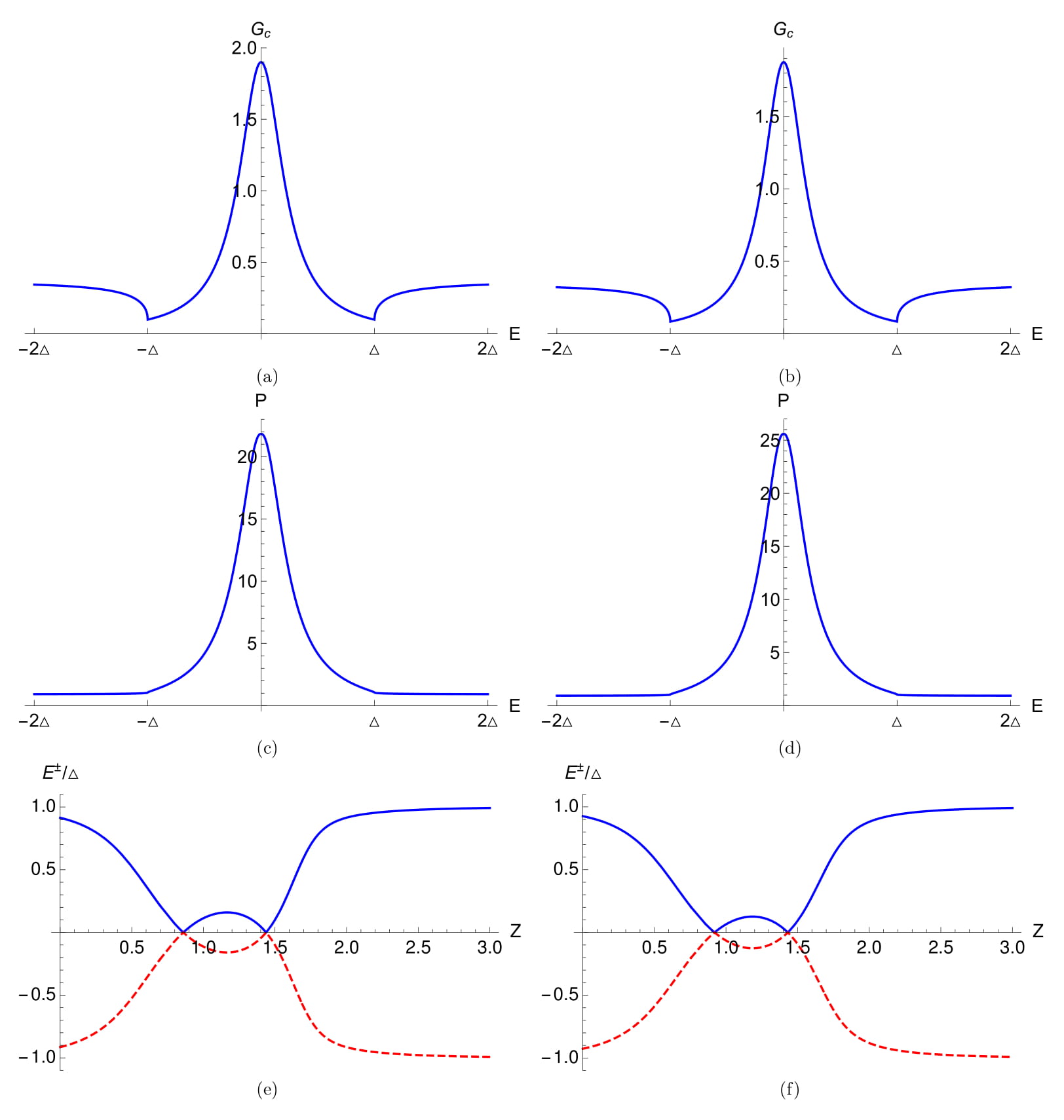}
\caption{\small \sl (a) Charge conductance vs energy with $S=21/2$, $m'=-1/2$, $Z=0.86$, $J=0.4$, $k_{F}a=0.8437\pi$, (b) Charge conductance vs energy with $S=23/2$, $m'=-1/2$, $Z=0.92$, $J=0.4$, $k_{F}a=0.8492\pi$, (c) Probability density vs energy with $S=21/2$, $m'=-1/2$, $Z=0.86$, $J=0.4$, $k_{F}a=0.8437\pi$, (d) Probability density vs energy with $S=23/2$, $m'=-1/2$, $Z=0.92$, $J=0.4$, $k_{F}a=0.8492\pi$, (e) Energy bound states as a function of interface transparency $Z$. Parameters are $S=21/2$, $m'=-1/2$, $J=0.4$, $k_{F}a=0.8437\pi$, (f) Energy bound states as a function of interface transparency $Z$. Parameters are $S=23/2$, $m'=-1/2$, $J=0.4$, $k_{F}a=0.8492\pi$. Here charge conductance is in units of $e^2/h$.}
\end{figure}
Next in Fig.~5 we plot the charge conductance and probability density as a function of energy $E$. We see that there is a peak at zero energy ($E=0$) in the conductance spectra due to the YSR bound states. We take large impurity spin for (a) $S=21/2$, (b) $S=23/2$ and small exchange interaction $J=0.4$. In Figs.~5(c), (d) zero energy peak is also observed in the probability density. For Figs.~5(c), (d) we take the same parameters as in Figs.~5(a), (b) respectively. {The calculated real part of poles of conductance in Figs.~5(a) and 5(b) are: $\pm 0.00012754\Delta$ and $\pm 0.00018198\Delta$ respectively and they match with the peak shown in Figs.~5(a) and 5(b). In Figs.~5(e) and (f) we plot energy bound states as a function of interface transparency ($Z$) for the same parameters as shown in Figs.~5(a) and (b) respectively. In Fig.~5(e) for $S=21/2$, we see that two bound state energies merge at $Z=0.86$ and $1.44$, where we see the zero energy peaks in the conductance spectra due to the YSR bound states. In Fig.~5(f) for $S=23/2$, we note that energy bound states merge at $Z=0.92$ and $1.43$, where zero energy peaks due to the YSR states are seen in the conductance spectra.}\par
Ref.~[\cite{pers}] also studied YSR states in the conductance spectra below the superconducting energy gap. The peak they see is due to the YSR states only near the gap edge. But in contrast, we also see the peak at zero energy $E=0$ due to YSR bound states.\par
\section*{Yu-Shiba-Rusinov bound states: Arbitrary junction length}
In the previous sections we mainly focus on short junction limit. In this limit we see that YSR bound states occur at $E=0$ in the conductance spectra for low values of $J$ and high $S$ values. In this section we study the effect of junction length on YSR bound states. We provide a comparison of YSR bound states between short and long junction.\par
For an electron with spin up incident, the wavefunction in the normal metal region I is given by for the long junction limit following Ref.~\cite{book},
\begin{equation}
\psi_{N}^{I}(x)=\begin{pmatrix}
                    u\\
                    0\\
                    0\\
                    0
                  \end{pmatrix}e^{ik_{e}x}\phi_{m'}^{S}+r_{ee}^{\uparrow\uparrow}\begin{pmatrix}
                  v\\
                  0\\
                  0\\
                  0
                 \end{pmatrix}e^{-ik_{e}x}\phi_{m'}^{S}+r_{ee}^{\uparrow\downarrow}\begin{pmatrix}
                 0\\
                -v\\
                 0\\
                 0
                \end{pmatrix}e^{-ik_{e}x}\phi_{m'+1}^{S}+r_{eh}^{\uparrow\uparrow}\begin{pmatrix}
                0\\
                0\\
               -v\\
                0
               \end{pmatrix}e^{ik_{h}x}\phi_{m'+1}^{S}+r_{eh}^{\uparrow\downarrow}\begin{pmatrix}
               0\\
               0\\
               0\\
               v
              \end{pmatrix}e^{ik_{h}x}\phi_{m'}^{S},\mbox{for $x<0,$}
\end{equation}              
Similarly the wavefunction in the normal metal region II is given by-
\begin{eqnarray}
\psi_{N}^{II}(x)=t_{ee}^{'\uparrow\uparrow}\begin{pmatrix}
                                     u\\
                                     0\\
                                     0\\
                                     0
                                     \end{pmatrix}e^{ik_{e}x}\phi_{m'}^{S}+t_{ee}^{'\uparrow\downarrow}\begin{pmatrix}
                                     0\\
                                     u\\
                                     0\\
                                     0
                                    \end{pmatrix}e^{ik_{e}x}\phi_{m'+1}^{S}+b_{ee}^{\uparrow\uparrow}\begin{pmatrix}
                                    v\\
                                    0\\
                                    0\\
                                    0
                                   \end{pmatrix}e^{-ik_{e}(x-a)}\phi_{m'}^{S}+b_{ee}^{\uparrow\downarrow}\begin{pmatrix}
                                   0\\
                                  -v\\
                                   0\\
                                   0
                                  \end{pmatrix}e^{-ik_{e}(x-a)}\phi_{m'+1}^{S}\nonumber\\+c_{eh}^{\uparrow\uparrow}\begin{pmatrix}
                                  0\\
                                  0\\
                                 -v\\
                                  0
                                 \end{pmatrix}e^{ik_{h}(x-a)}\phi_{m'+1}^{S}+c_{eh}^{\uparrow\downarrow}\begin{pmatrix}
                                 0\\
                                 0\\
                                 0\\
                                 v
                                \end{pmatrix}e^{ik_{h}(x-a)}\phi_{m'}^{S}+a_{eh}^{\uparrow\uparrow}\begin{pmatrix}
                                0\\
                                0\\
                                u\\
                                0
                               \end{pmatrix}e^{-ik_{h}x}\phi_{m'+1}^{S}+a_{eh}^{\uparrow\downarrow}\begin{pmatrix}
                               0\\
                               0\\
                               0\\
                               u
                              \end{pmatrix}e^{-ik_{h}x}\phi_{m'}^{S},\mbox{for $0<x<a,$}
                              \end{eqnarray}
The corresponding wavefunction for the superconductor is-
\begin{equation}
\psi_{S}(x)=t_{ee}^{\uparrow\uparrow}\begin{pmatrix}
                              u\\
                              0\\
                              0\\
                              v
                             \end{pmatrix}e^{iq_{+}x}\phi_{m'}^{S}+t_{ee}^{\uparrow\downarrow}\begin{pmatrix}
                             0\\
                             u\\
                             -v\\
                             0
                             \end{pmatrix}e^{iq_{+}x}\phi_{m'+1}^{S}+t_{eh}^{\uparrow\uparrow}\begin{pmatrix}
                             0\\
                             -v\\
                             u\\
                             0
                             \end{pmatrix}e^{-iq_{-}x}\phi_{m'+1}^{S}+t_{eh}^{\uparrow\downarrow}\begin{pmatrix}
                             v\\
                             0\\
                             0\\
                             u
                             \end{pmatrix}e^{-iq_{-}x}\phi_{m'}^{S},\mbox{for $x>a.$}
\end{equation}                             
For $\rvert E \rvert<<E_{F}$, we can write $k_{e,h}\approx k_{F}\pm \frac{E}{2\Delta\xi}$, where $\xi=E_{F}/(k_{F}\Delta)$ is the Cooper pair coherence length\cite{Kri}. The boundary conditions at different interfaces of our system are mentioned before in Eqs.~6,7,8,9. By imposing the boundary conditions on the wavefunctions mentioned in Eqs.~16,17,18 one can get the different scattering amplitudes. After getting the scattering amplitudes, using Eq.~10 we can calculate the charge conductance for arbitrary junction length. In Fig.~6 we plot the charge conductance as a function of energy $E$ for different junction length $a$. In Fig.~6(a) we concentrate on the short junction limit ($a<\xi$). In this limit we take three different values of $a$ ($a=0$, $a=0.1\xi$, $a=0.4\xi$). We see that a peak appears at zero energy ($E=0$) in the conductance spectra due to the YSR states. We also see that peaks in conductance formed due to the merger of the YSR bound states at $E=0$ are robust { to change in length $a$ of the junction.} We take large impurity spin $S=21/2$, and small exchange interaction $J=0.4$. However, these YSR peaks at $E=0$ are unfortunately not as robust to changes in other parameters, e.g., $Z, J, S, m'$ 
\begin{figure}[h]  
\includegraphics[width=.99\textwidth]{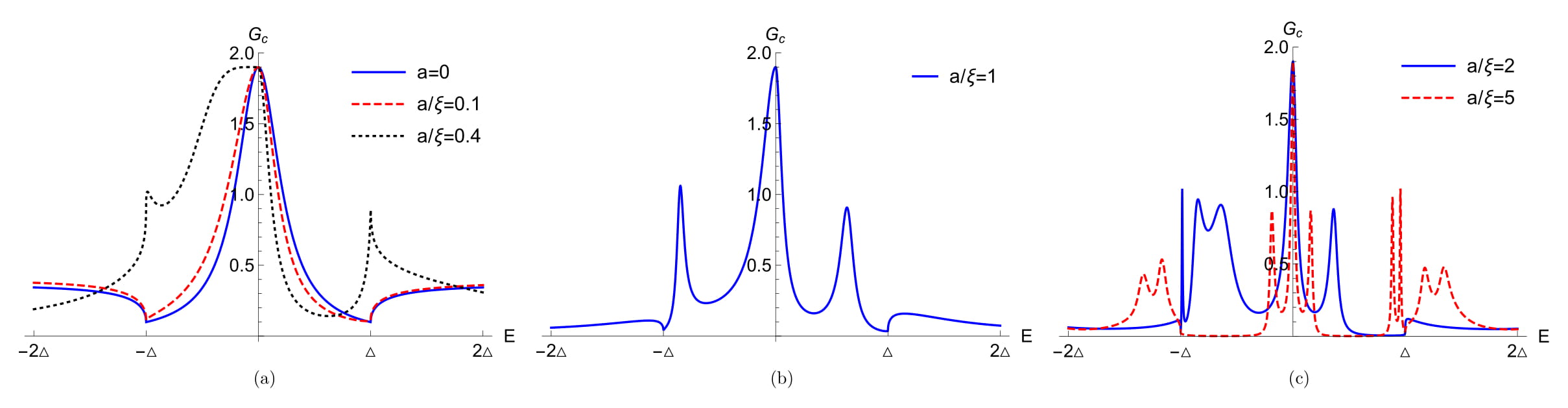}
\caption{\small \sl Charge conductance as a function of energy for different junction length ($a$). Parameters are $S=21/2$, $m'=-1/2$, $Z=0.86$ , $J=0.4$, $k_{F}a=0.8437\pi$. Here charge conductance is in units of $e^2/h$.}
\end{figure}
Next, in Fig.~6(b) we concentrate on the intermediate junction limit ($a\sim\xi$).  
 { In this limit we also see a peak at $E=0$ in the conductance spectra due to the YSR bound states. Further, we see peaks due to YSR bound states, present asymmetrically at both positive and negative energies near the gap edge within the energy gap.} Finally, in Fig.~6(c) we focus on long junction limit ($a>\xi$). In this limit we take two different values of $a$ ($a=2\xi$, $a=5\xi$). {We see that there is a peak at $E=0$ in the conductance spectra. Further, many YSR peaks appear in the subgap regime of the conductance spectra.} Similar features have also been seen in Ref.~\cite{ben}, where magnetic impurity is outside the superconductor. Many YSR peaks appear below the gap in the conductance spectra with increase of temperature (see Figs.~4(b),(c) of Ref.~\cite{ben}). When temperature is increased, superconducting coherence length decreases and the junction behaves as a long junction ($a>\xi$). Thus, in the long junction limit many YSR peaks are seen in the subgap regime of the conductance spectra similar to our work.\par  
\section*{Conclusion}
In conclusion, we have studied the YSR bound states in the vicinity of a s-wave superconductor in presence of a magnetic impurity and probably this is the first time YSR states have been analyzed using BTK approach. We mainly focus on the zero energy in the conductance spectra. We see that when HSM does not flip there is a dip at zero energy in the conductance spectra, but for spin flip case a zero energy peak is observed due to the YSR bound states. {We plot the real part of the complex poles of conductance as a function of interface transparency $Z$ and see that two YSR bound states merge at particular values of $Z$. Where the bound states merge, gives a zero energy peak in conductance spectra.} We also study the effect of arbitrary junction length on YSR bound states. We see that YSR peaks appear at $E=0$ in the conductance spectra for {any arbitrary length of the junction}. Further, for long junction many YSR peaks are seen in the subgap regime of the conductance spectra. {This $E=0$ peak is robust to change in junction length $a$, but not to change in other parameter values like for  exchange interaction ($J$), interface transparency ($Z$), etc.} The YSR bound state at $E=0$, is however non-topological in contrast to the topological $E=0$ bound state seen in Normal metal-insulator-p wave superconductor junction\cite{setiawan}.  

\section*{Acknowledgments}
This work was supported by the grant ``Non-local correlations in nanoscale systems: Role of decoherence, interactions, disorder and pairing symmetry'' from SCIENCE \& ENGINEERING RESEARCH BOARD, New Delhi, Government of India, Grant No.  EMR/20l5/001836,
Principal Investigator: Dr. Colin Benjamin, National Institute of Science Education and Research, Bhubaneswar, India.
\section*{Author contributions statement}
C.B. conceived the proposal,  S.P. did the calculations on the advice of C.B., C.B. and S.P. analyzed the results and wrote the paper.  Both  authors reviewed the manuscript. 
\section*{Competing interests}
The authors declare no competing interests.

\end{document}